\documentclass[mathleft
]{an}
\usepackage{graphicx}
\usepackage{times}
\usepackage{natbib}

\bibpunct{(}{)}{;}{a}{}{,}
\usepackage[pdfpagelabels]{hyperref} 	
\begin{document}

\Pagespan{907}{}
\Yearpublication{2011}%
\Yearsubmission{2011}%
\Month{12}%
\Volume{332}%
\Issue{9/10}%
\DOI{10.1002/asna.20111623}%

\def\cf{{\it cf. }}
\def\eg{{\it e.g. }}
\def\ie{{\it i.e. }}
\def\etal{{\it et al. }}
\def\Pcyc{P_{\rm cyc}}
\def\Prot{P_{\rm rot}}
\def\cms{\rm ~cm~s^{-1}}
\def\ms{\rm ~m~s^{-1}}
\def\acos{\rm acos}

\newcommand{\derive}[2]{{\frac{\mathrm{d}#1}{\mathrm{d}#2}}}
\newcommand{\derivep}[2]{{\frac{\partial#1}{\partial#2}}}
\newcommand{\erf}[1]{{\rm erf \left(#1\right)}}

\title{Effects of turbulent pumping on stellar activity cycles}
\author{O. Do Cao \and A. S. Brun}

\titlerunning{Effects of turbulent pumping on stellar activity cycles}
\authorrunning{Do Cao, Brun}
\institute{Laboratoire AIM Paris-Saclay, CEA/Irfu Universit\'e Paris-Diderot CNRS/INSU, 91191 Gif-sur-Yvette, France}
\received{2011 Oct 5}
\accepted{2011 Oct 27}
\publonline{2011 Dec 12}

\abstract{Stellar magnetic activity of solar like stars is thought to be due to an internal dynamo. While the Sun has been the subject of intense research for refining dynamo models, observations of magnetic cyclic activity in solar type stars have become more and more available, opening a new path to understand the underlying physics behind stellar cycles. For instance, it is key to understand how stellar rotation rate influences magnetic cycle period $\Pcyc$. Recent numerical simulations of advection-dominated Babcock Leighton models have demonstrated that it is difficult to explain this observed trend given a) the strong influence of the cycle period to the meridional circulation amplitude and b) the fact that 3D models indicate that meridional flows become weaker as the rotation rate increases. In this paper, we introduce the turbulent pumping mechanism as another advective process capable also of transporting the magnetic fields. We found that this model is now able to reproduce the observations under the assumption that this effect increases as $\Omega^2$.  The turbulent pumping becomes indeed another major player able to circumvent the meridional circulation. However, for high rotation rates ($\Omega \simeq 5 \Omega_\odot$), its effects dominate those of the meridional circulation, entering a new class of regime dominated by the advection of turbulent pumping and thus leading to a cyclic activity qualitatively different from that of the sun.}

\maketitle

\section{Introduction}
\subsection{Magnetic cycles in Sun and Stars}
The Sun exhibits a magnetic period of about 22 years. Observations based on the surface activity seen in Ca II-HK emission lines in the infrared and X-ray flux \citep{2009ASPC..416..375S} revealed the existence of such cycles with comparable periods (from 5 to 25 years) in solar like stars, \ie constituted of a deep convective envelope and a radiative interior. The availability of more and more data on the magnetism of G\&K stars provide us a new insight in understanding the global scale generation of their magnetic fields. This magnetism and regular activity is thought to be due to a magnetic dynamo operating in the bulk and at the edges of the convection zone \citep{2010A&A...509A..32J,1996ApJ...460..848B,1999ApJ...524..295S,2004SoPh..224..161N}.

In order to complement costly full 3D magnetohydrodynamical (MHD) simulations, a useful approach has been to make use of the mean field dynamo theory \citet{moffatt1978}. This method has the advantage that it only deals with the large scale magnetic field, assuming some parametrization of the underlying small scale turbulence and magnetism. In these theories, the toroidal field is generally assumed to be generated at the base of the convection zone (BCZ) in a region called the tachocline, where both a radial and latitudinal shear act. What is more poorly known though is the source of the poloidal field to close the dynamo loop (\ie $B_{pol} \rightarrow B_{tor} \rightarrow B_{pol}$). Among the various existing mechanisms, one of the most promising one is the Babcock Leighton (BL) mechanism first proposed by \citet{babcock1961} and further elaborated by \citet{1969ApJ...156....1L}. In BL models, the poloidal field is generated at the surface by the transport and decay of tilted bipolar magnetic regions which are formed by twisted buoyant magnetic flux ropes developed at the tachocline \citep{browning2006,jouve2009}. Synoptic magnetographic monitoring over solar cycles 21 and 22 has offered strong evidence for such dynamo action \citep{1989Sci...245..712W,1991ApJ...383..431W,2010A&amp;A...518A...7D}. The BL flux transport dynamo models have thus been recently the favored one and have demonstrated some success at reproducing solar observations assuming either an advection \citep[e.g.][]{1999ApJ...518..508D} or a diffusion dominated \citep[e.g.][]{2008ApJ...673..544Y} regime for the transport of field from the surface down to the tachocline. Alternative $\alpha$ effects such as helical turbulence may also be used in mean field dynamo models \citep{2010LRSP....7....3C} but we will not do so here.

In the past decades, dynamo action has been studied preferentially on the Sun, but some works have addressed the question of the applicability of such models to other solar type stars which possess different rotation rates and activity levels \citep{1996ApJ...460..848B,1984ApJ...287..769N}. One difficulty of such observational programs is that they require long term observations since stellar cycle periods are likely to be commensurate to the solar 11-yr sunspots cycle period (or 22-yr for a full cycle including two polarity reversals of the global poloidal field). The biggest survey to date is from Mt. Wilson \citep{2009ASPC..416..375S}, but gathered data from the literature are now available for solar type stars with enough statistics \citep{2011arXiv1109.4634W}. The systematic analysis of these data revealed that the cycle period is shorter as the star rotates faster and that above a certain rotation rate the X-ray luminosity saturates (see below).  The pioneering work of \citet{1984ApJ...287..769N} found that $P_{\rm cyc} \propto P_{\rm rot}^n$, with $n = 1.25 \pm 0.5$. However, \citet{1998ApJ...498L..51B} argued that an "Active'' (with $n = 0.8$ ) and "Inactive'' (with $n = 1.15$) branches, segregating respectively Young and old stars, could actually coexist. The Sun is found to actually lie in between. These scalings take into account the existence of a primary (Hale) cycle and a secondary (Gleissberg) cycle.  However, saturation of the X-ray luminosity limits this scaling to moderate rotation rates \citep{2003A&amp;A...397..147P,2011arXiv1109.4634W}. For G type stars this saturation is found for rotation rate above 35 ${\rm km~s^{-1}}$, for K type stars at about 10 ${\rm km~s^{-1}}$ and for M dwarfs around 3-4 ${\rm km~s^{-1}}$ \citep{2008ApJ...676.1262B,2009ApJ...692..538R}. How stellar magnetic flux scales with rotation rate is thus also important to understand since it tells us how the magnetic field generated by dynamo action inside the stars emerges and imprints the stellar surface \citep{2008JPhCS.118a2032R} and if it actually saturates.

In the framework of BL flux transport models including a meridional circulation (MC), the $\Pcyc-\Prot$ relationship can be reproduced only if the meridional flow is proportional to the rotation rate of the star \citep{2004SoPh..224..161N,2001ASPC..248..235D,2001ASPC..248..189C,2007AdSpR..40..891N}. However recent theoretical work by \citet{2007ApJ...669.1190B} and \citet{brown2008} indicate instead that the amplitude of the meridional flow weakens as the rotation rate is increased. The recent work of \citet{2010A&A...509A..32J} (hereafter JBB2010) shows indeed that with such scaling, they cannot recover the observational trend of a shorter cycle period for faster rotation rates, unless a multicellular MC flow is assumed. We wish here instead to explore the influence of the turbulent pumping using a simple unicellular MC without invoking a more complex pattern.

\subsection{Turbulent pumping in stellar dynamo}\label{section_turbulentpumpingintro}
Magnetic pumping refers to transport of magnetic fields in convective layers that does not result from bulk motion. One particular case is turbulent pumping. In inhomogeneous convection due to density stratification, convection cells take the form of broad hot upflows surrounded by a network of downflow lanes \citep{1991MNRAS.253..479C,2008ApJ...673..557M}. In such radially asymmetric convection, numerical simulations show that the magnetic field is preferentially dragged downward \citep{2001ApJ...549.1183T}. This effect has been demonstrated to operate in the bulk of the solar convection zone. A significant equatorward latitudinal component also arise when rotation becomes important, \ie when the Rossby number is less than unity. Turbulent pumping speeds of a few$\ms$ can be reached according to the numerical simulations of \citet{2006A&amp;A...455..401K}. Therefore, its effects are expected to be comparable to those of meridional circulation.

In spite of those results, the effects of turbulent pumping rarely have been considered in mean field models. As the latter were able to reproduce rather well the large scale magnetic field using only the alpha effect and differential rotation to drive the dynamo, the pumping effect was thought to be an unnecessary complication.

A first approach showing the importance of pumping in the solar cycle was made by \citet{1992ASPC...27..536B}. Since then, turbulent pumping has been a useful approach to tackle the problem of storage of magnetic fields. Indeed, for magnetic flux ropes to be buoyantly unstable and to emerge at the surface to form bipolar magnetic region with the appropriate tilt, numerical simulations have shown that their strength must be as high as  $10^4 - 10^5 \rm ~G$ \citep{2007AN....328.1104J,2004ASPC..325...47F,1987ApJ...316..788C}. One important limitation of this scenario is that $10^5 \rm ~G$ represents a magnetic energy density higher by an order of magnitude than the kinetic energy density, \ie one needs to create super equipartition magnetic structures. Therefore, a stable layer is required to store and amplify the magnetic fields. For this process to occur, differential rotation must be able to develop intense toroidal magnetic fields either within the tachocline or in the convection envelope. Pumping could be the transport needed to get the poloidal field down to the tachocline and could also maintain strongly buoyant structures from rising, thus helping them to become even stronger.

More interestingly in the regard of our work, turbulent pumping have shown remarkable properties regarding stellar cycles. \citet{2008A&A...485..267G} (hereafter GdG2008) have demonstrated that the magnetic cycle period $\Pcyc$ is no longer dominated by the meridional flow speed $v_0$ but instead by the radial turbulent pumping $\gamma_r$ following the relation :

\begin{equation}
P_{\rm cyc} \propto v_0^{-0.12} \gamma_r^{-0.51} \gamma_\theta^{-0.05}
\end{equation}

\noindent available in the range $v_0=[500;3000]$, $\gamma_{r0}=[20;120]$, $\gamma_{\theta 0}=[60;140]$ in $\rm cm ~s^{-1}$. Another interesting feature of this model is that the surface magnetic field no longer shows the strong concentration in the polar region that usually characterizes Babcock Leighton dynamo solutions operating in the advection dominated regime. This can be traced primarily to the efficient downward turbulent pumping that subducts the poloidal field as it is carried poleward by the meridional flow.

With the STELEM code, we aim to study the effect of turbulent pumping on $\Pcyc$ and to see under which conditions the $\Pcyc-\Prot$ relationship can be recovered. The paper is organised in the following manner. In Section \ref{section_model}, we describe the equations, the initial and boundary conditions, the ingredients of the model and the standard case in which the turbulent pumping is \emph{not} included. Section \ref{section_pumping} shows how the dynamo model behaves in the presence of turbulent pumping and we conclude in Section \ref{section_conclusion}.

\section{The model}\label{section_model}
\subsection{The model equation}

To model the stellar global dynamo operating in solar like stars, we start from the hydromagnetic induction equation, governing the evolution of the magnetic field $\bf{B}$ in response to advection by a flow field $\bf{V}$ acting against the magnetic dissipation characterised by the molecular magnetic diffusivity $\eta_m$ :

\begin{equation}
  \derivep{\bf B}{t}=\nabla\times ({\bf V} \times{\bf B})-\nabla\times(\eta_m\nabla\times{\bf B})
\end{equation}

 As we are working in the framework of mean field theory, we are interested in the large-scale magnetic field on time scales longer than the turbulent time scale. We express both magnetic and velocity fields as a sum of a mean component (usually defined as a longitudinal average) and small-scale fluctuating component. For instance, the magnetic field is decomposed as :

\begin{equation}
  \bf B = \langle \bf B \rangle + \bf b
\end{equation}

Upon this separation and averaging procedure, the induction equation for the mean component becomes

\begin{eqnarray}
\derivep{\bf \langle B\rangle}{t}&=&\nabla\times ({\bf \langle V\rangle} \times{\bf \langle B\rangle})
+\nabla\times\langle{\bf v}\times {\bf b}\rangle \nonumber \\
&-&\nabla\times(\eta_m\nabla\times{\bf \langle B\rangle})
\end{eqnarray}

We drop now the averaging symbol $\langle \rangle$ for the sake of clarity in the rest of the paper. A closure relation must then be used to express the mean electromotive force (emf) $\bf \mathcal{E}  = \bf \langle{ v}\times { b}\rangle$  in terms of mean magnetic field, leading to a simplified mean-field equation. If the mean magnetic field varies slowly in time and space, the emf can be represented in terms of $\bf B$ and its gradients

\begin{equation}
  \mathcal{E}_i = a_{ij} B_j + b_{ijk} \derivep{B_j}{x_k} + ...
\end{equation}

where $a_{ij}$ and $b_{ijk}$ are in the general case tensors containing the transport coefficients, and the dots indicate that higher order derivatives can be taken into account. Summation over repeated indices is assumed. The tensors $a_{ij}$ and $b_{ijk}$ cannot, in general, be expressed from first principles due to the lack of a comprehensive theory of convective turbulence. In the kinematic regime where the magnetic energy is negligible in comparison to the kinetic energy, the most simple approximation is to neglect all correlations higher than second order in the fluctuations. This is the so-called first order smoothing approximation (FOSA). In the previous works \citep[see][for a recent review]{2010LRSP....7....3C}, they consider the simple case of isotropic turbulence, the tensor $a_{ij}$ reduces into a single scalar giving rise to the $\alpha$-effect. However, we consider in this paper the full tensor non-isotropic case for $a$. The emf can then in general be written as

\begin{equation}
  \bf \mathcal{E} = (\alpha \bf B + \bf \gamma \times \bf B) - \beta \bf \nabla \times \bf B
\end{equation}

where $\alpha$ is a scalar referring to the standard $\alpha$-effect. As we work with BL models, we assume instead the poloidal field to be generated at the surface, so that we will replace the $\alpha \bf B$ term by an non local source term $S$ (details are described below). The term $\gamma$ is the turbulent pumping and $\beta$ is defined such that $b_{ijk}=\beta\epsilon_{ijk}$ (with $\epsilon_{ijk}$ is the Levi-Civita tensor).

As $\gamma$ and $\beta$ originates from the same velocity field, they both in principle depend on $\bf v$ \citep[see][]{krause1980mean}. In complex configurations as rotating convective spherical shells, it is not clear though how they are related in spite of numerous efforts in determining precisely the turbulent transport coefficients \citep[\eg][]{2009A&amp;A...500..633K,1990A&amp;A...232..277B,2001A&amp;A...376..713O,2006JFM...553..401C}. In the models computed in this work, we thus do not consider any relationship between these two quantities. As can be seen in equations \ref{eqA2} and \ref{eqB2}, $\beta$ can be directly interpreted as an effective diffusion coefficient whose effects are known. Hence, we have focused on the effects of varying $\gamma$ for a fixed $\beta$ even if a full parameter study in $\{\gamma;\beta\}$ space should in principle be done.

Working in spherical coordinates and under the assumption of axisymmetry, we write the total mean magnetic field {\bf B}, velocity field {\bf V} and the turbulent pumping (where we neglect $\gamma_\phi$ because its amplitude is much less than the differential rotation) as :

\begin{eqnarray}
{{\bf B}}(r,\theta,t)&=&\nabla\times (A_{\phi}(r,\theta,t) \hat {\bf e}_{\phi})+B_{\phi}(r,\theta,t) \hat {\bf e}_{\phi} \\
{{\bf V}}(r,\theta)&=&{\bf v_{p}}(r,\theta) + r\sin\theta \, \Omega(r,\theta) \hat {\bf e}_{\phi}, \\
{{\bf \gamma}}(r,\theta)&=&{\bf \gamma_{p}}(r,\theta) = {\bf \gamma_r} (r,\theta) \hat{\bf e}_r + {\bf \gamma_\theta}(r,\theta) \hat{\bf e}_\theta
\end{eqnarray}

\noindent where $\bf v_p$ is the poloidal velocity field. Reintroducing this poloidal/toroidal decomposition of the field in the mean induction equation, we get two coupled partial differential equations, one involving the vector potential $A_{\phi}$ and the other concerning the toroidal field $B_{\phi}$. The corresponding dimensionless equations are then :

\begin{eqnarray}
\label{eqA2}
\derivep{A_{\phi}}{t}&=&\frac{\eta}{\eta_{t}} \left(\nabla^{2}-\frac{1}{\varpi^{2}}\right) A_{\phi}+C_{s}S(r,\theta,B_{\phi}) \nonumber\\
&&- R_{e}\frac{\bf{v}_{p}}{\varpi}\cdot\nabla(\varpi A_{\phi}) \nonumber\\
&&+ \frac{1}{\varpi}\left(C_{\gamma r} \gamma_r \hat{\bf e}_r + C_{\gamma\theta} \gamma_\theta \hat{\bf e}_\theta\right)\cdot\nabla(\varpi A_{\phi})
\end{eqnarray}

\begin{eqnarray}
\label{eqB2}
\frac{\partial {B_{\phi}}}{\partial t}&=&\frac{\eta}{\eta_{t}} (\nabla^{2}-\frac{1}{\varpi^{2}})B_{\phi}
+\frac{1}{\varpi}\derivep{(\varpi B_{\phi})}{r}\derivep{(\eta/\eta_{t})}{r} \nonumber  \\
&-&R_{e}\varpi {\bf v}_{p}\cdot\nabla(\frac{B_{\phi}}{\varpi})-R_{e}B_{\phi}\nabla\cdot{\bf v}_{p} \nonumber  \\
&-&\varpi \left(C_{\gamma r} \gamma_r \hat{\bf e}_r + C_{\gamma\theta} \gamma_\theta \hat{\bf e}_\theta\right)\cdot\nabla \left(\frac{B_{\phi}}{\varpi}\right) \nonumber  \\
&-&B_{\phi}\nabla\cdot\left(C_{\gamma r} \gamma_r \hat{\bf e}_r + C_{\gamma\theta} \gamma_\theta \hat{\bf e}_\theta\right) \nonumber \\
&+&C_{\Omega}\varpi(\nabla\times(\varpi A_{\phi}{\bf \hat{e}}_{\phi}))\cdot\nabla\Omega
\end{eqnarray}

where $\varpi=r\sin\theta$, $\eta=\eta_m+\beta$, $\eta_{t}$ is the turbulent magnetic diffusivity (diffusivity in the convective zone) and $S(r,\theta,B_{\phi})$ the Babcock-Leighton surface source term for poloidal field (we neglect its contribution in the generation of the toroidal field compare to the shear applied by the differential rotation. We define the diffusive timescale as $\tau_\eta = R_\odot^2/\eta_t$. Note that our velocity field is time-independent since we will not assume any fluctuations in time of the differential rotation $ \Omega$ or of the meridional circulation ${\bf v_{p}}$.

In order to write these equations in a dimensionless form, we choose as length scale the solar radius $R_{\odot}$ and as timescale the diffusion time $R_{\odot}^2/\eta_{t}$ based on the envelope diffusivity $\eta_{t} = 5 ~10^{10}\,\rm cm^2\rm s^{-1}$. This leads to the appearance of five control parameters $C_{\Omega}=\Omega_{0}R_{\odot}^2/\eta_{t}$, $C_{s}=s_{0}R_{\odot}/\eta_{t}$ and $R_{e}=v_{0}R_{\odot}/\eta_{t}$ where $\Omega_{0}, s_{0}, v_{0}$ are respectively the rotation rate at the equator and the maximal amplitude of the surface source term and of the meridional flow. For the solar rotation rate, we have $\Omega_0/2\pi=456 \rm ~nHz$. We can also define similar dimensionless numbers $C_{\gamma r} = \gamma_{r0}R_\odot/\eta_t$ and $C_{\gamma\theta}=\gamma_{\theta0}R_\odot/\eta_t$ for turbulent pumping. As we don't know how the $C_\gamma$'s are related, we allow for any relation between them.

Equations \ref{eqA2} and \ref{eqB2} are solved with the STELEM code (Emonet \& Charbonneau 1998, private communication) in an annular meridional plane with the colatitude $\theta$ $\in [0,\pi]$ and the radius (in dimensionless units) $r \in [0.6,1]$ i.e. from slightly below the tachocline ($r=0.7$) up to the solar surface. The STELEM code has been thoroughly tested and validated thanks to an international mean field dynamo benchmark involving 8 different codes \citep{2008A&amp;A...483..949J}. At $\theta=0$ and $\theta=\pi$ boundaries, both $A_{\phi}$ and $B_{\phi}$ are set to 0. Both $A_{\phi}$ and $B_{\phi}$ are set to $0$ at $r=0.6$. At the upper boundary,  we smoothly match our solution to an external potential field, i.e. we have vacuum for $r \geq 1$ \citep[see][for more details]{2007A&amp;A...474..239J}. As initial conditions we are setting a confined dipolar field configuration, i.e. the poloidal field is set to $\sin\theta / r^{2}$ in the convective zone and to $0$ below the tachocline whereas the toroidal field is set to $0$ everywhere. All simulations have been carried with a resolution of 129x129 and the parameters used for the models are summarised in Table \ref{table_parameters}.

\subsection{The physical ingredients}\label{section_ingredients}

All quantities are in dimensionless values, and all the profiles are normalised to unity such that the dimensionless parameters set the strength of each ingredient.

\begin{figure}[!h]
  \includegraphics[width=3.5cm]{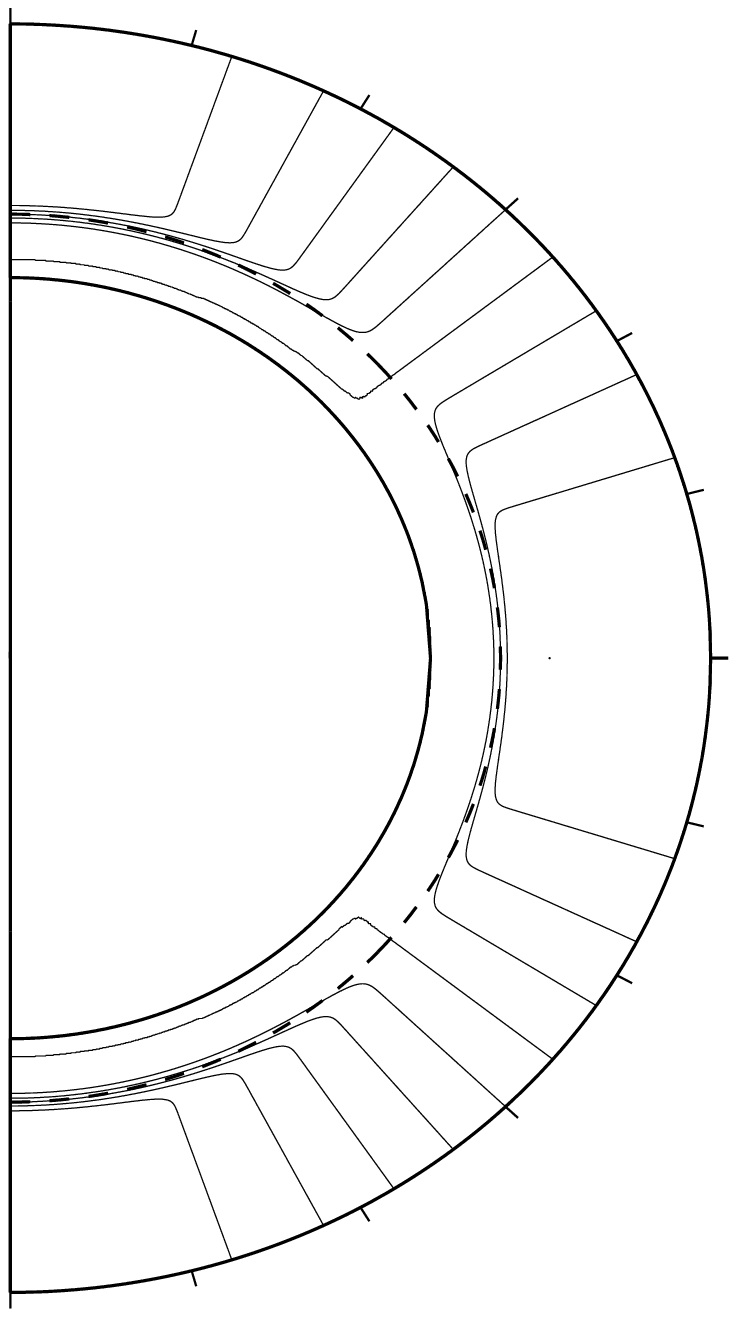} \hfill
  \includegraphics[width=3.5cm]{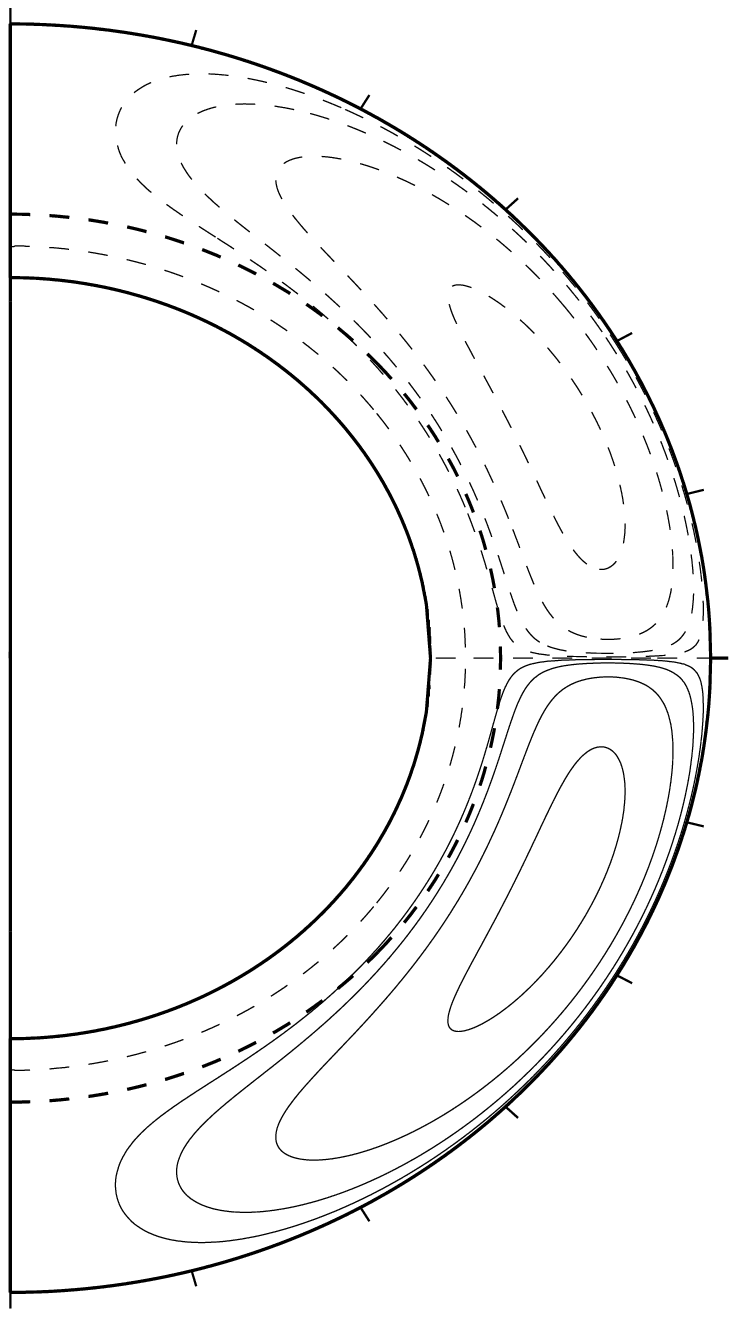}
  \caption{\textit{Left panel :} Isorotation lines. Contours are regularly spaced by $0.02$ and spans from $0.8$ to $1.0$. \textit{Right panel :} Meridional streamfunction. Contours are logarithmically spaced by $10^{0.5}$. Positive (negative) values are in solid (dash) line. The thick dash line locates the tachocline position.}
  \label{rotation}
\end{figure}

The rotation profile captures some realistic aspects of the Sun's angular velocity, deduced from helioseismic inversions \citep{2003ARA&A..41..599T}, assuming a solid rotation below 0.66 and a differential rotation above the interface (\cf Fig. \ref{rotation}).

\begin{eqnarray}
  \Omega(r,\theta)&=&\Omega_c + \frac{1}{2}\left[1+\rm erf\left(\frac{r-r_c}{d_1}\right)\right]  \nonumber \\
  &&\times \left[1 -\Omega_c-c_2\cos^2(\theta)\right]
\end{eqnarray}

\noindent with $r_c = 0.7$, $d_1 = 0.02$, $\Omega_c = 0.92$ and $c_2 = 0.2$. With this profile, the radial shear is maximal at the tachocline.

 We assume that the diffusivity in the envelope $\eta$ is dominated by its turbulent contribution $\eta_T$ in the convection zone  whereas in the stable interior, $\eta_c \ll \eta_T$. We smoothly match the two different constant values with an error function which enables us to quickly and continuously transit from $\eta_c$ to $\eta_T$ in the vicinity of the core-envelope interface i.e.

\begin{equation}
  \eta(r) = \frac{\eta_c}{\eta_t} + \frac{1}{2}\left(1-\frac{\eta_c}{\eta_T} \right) \left[1+\rm \erf{\frac{r-r_c}{d_1}}\right]
\end{equation}

\noindent with $\eta_c=5 ~10^8 \rm cm^2s^{-1}$.

In Babcock-Leighton (BL) flux-transport dynamo models, the poloidal field owes its origin to the tilt of magnetic loops emerging at the solar surface. Thus, the source has to be confined to a thin layer just below the surface and since the process is fundamentally non-local, the source term depends on the variation of $B_{\phi}$ at the BCZ. A quenching term is introduced to prevent the magnetic energy from growing exponentially without bound (see Section \ref{section_conclusion}). The expression is then

\begin{eqnarray}
S(r,\theta,B_{\phi})&=& \frac{1}{2}\left[1+\erf{\frac{r-r_2}{d_2}}\right]\nonumber \\
&\times&\left[ 1-\erf{\frac{r-1}{d_2}}\right] \nonumber \\
&\times&\left[1+\left({\frac{B_{\phi}(r_{c},\theta,t)}{B_{0}}}\right)^{2}\right]^{-1} \\
&\times&\cos\theta \sin\theta B_{\phi}(r_{c},\theta,t) \nonumber
\end{eqnarray}

\noindent where $r_{2}=0.95$ is the location of the max of the BL source term, $d_{2}=0.01$ is the thickness of this layer and $B_{0}=10^5 \, \rm G$.

\begin{figure}[!h]
  \includegraphics[width=8cm]{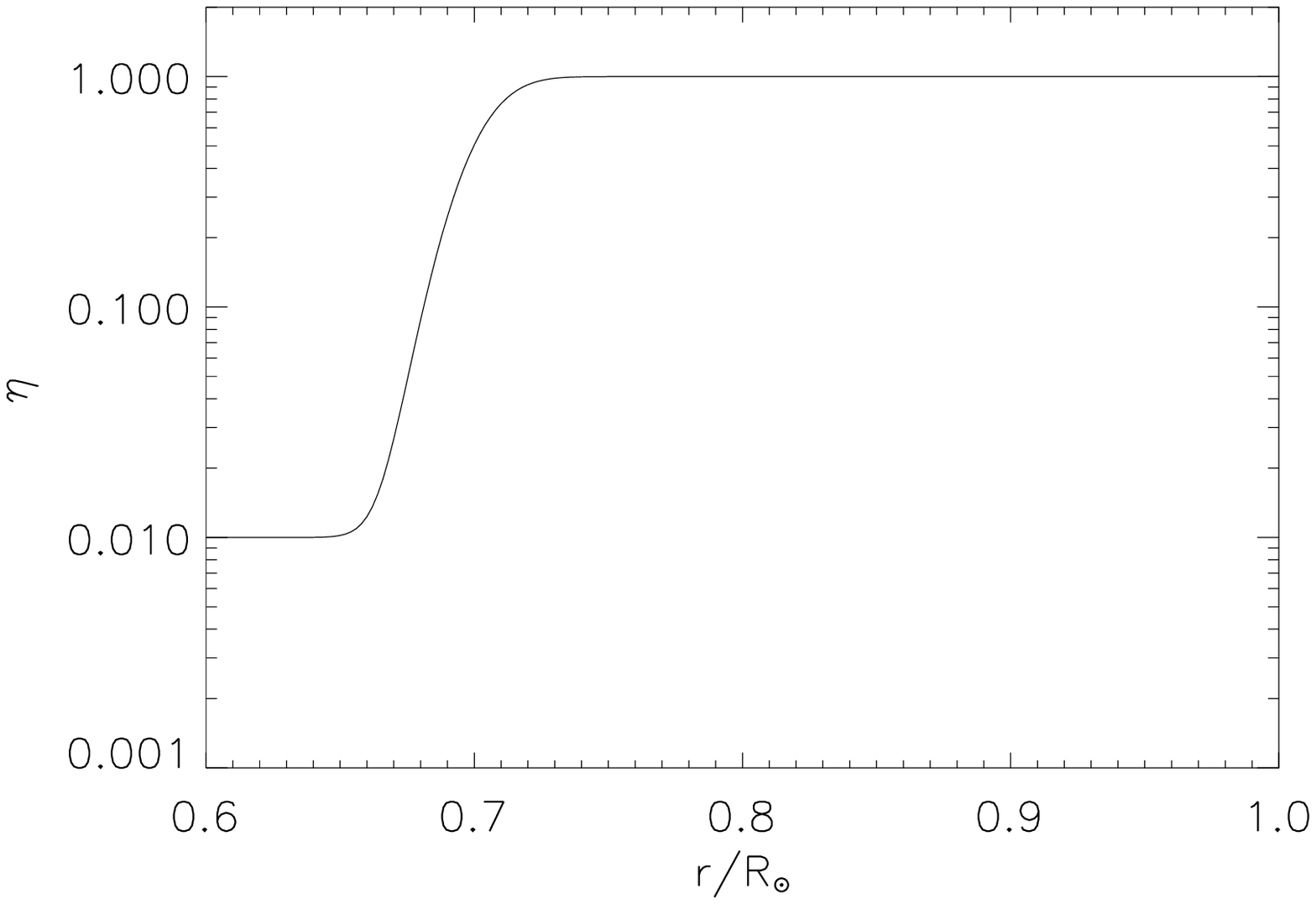}
  \includegraphics[width=8cm]{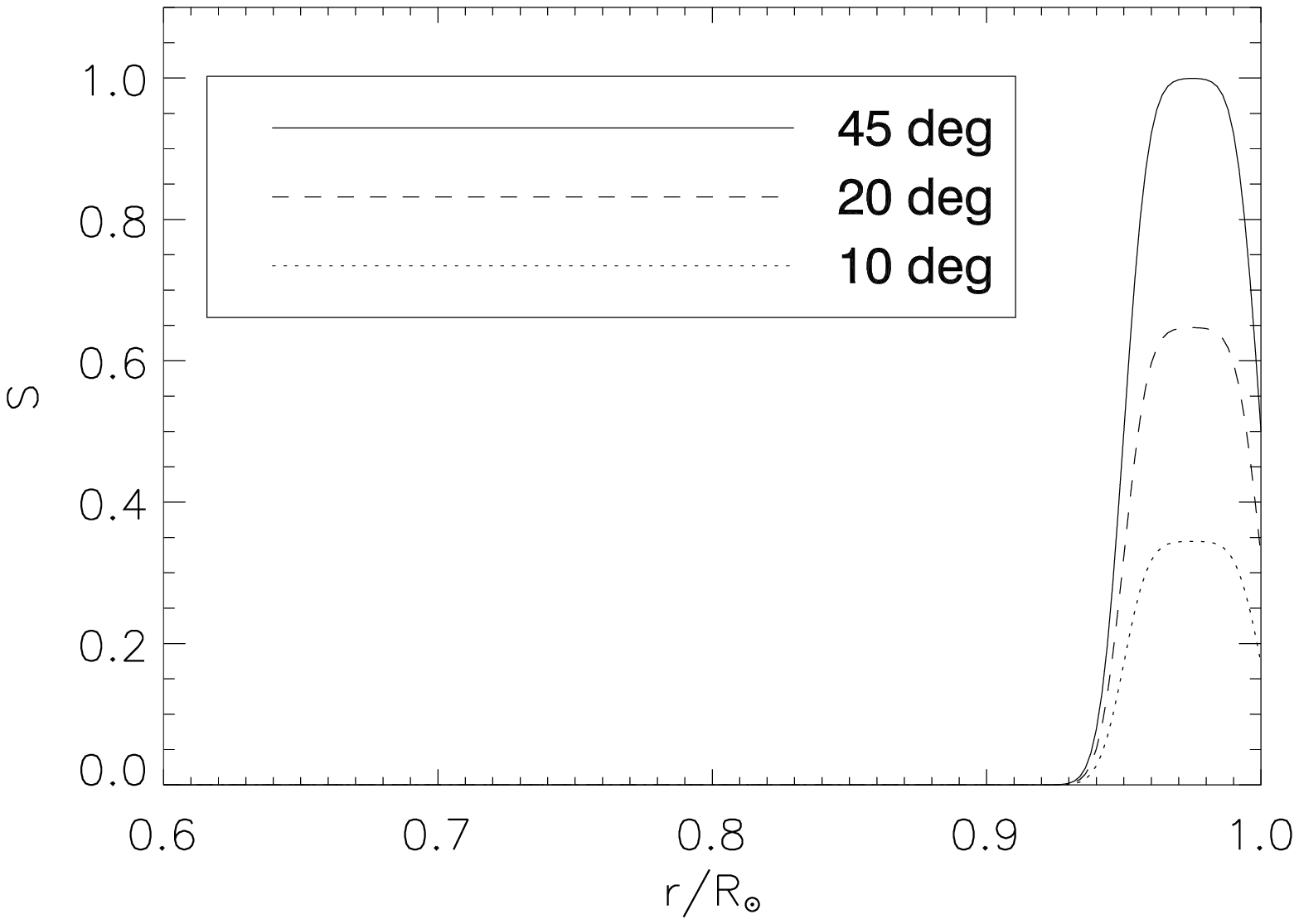}
  \caption{\textit{Upper panel :} Diffusivity profile as function of radius. \textit{Lower panel :} Babcock Leighton source term as function of radius for different colatitudes.}
  \label{profiles_physique}
\end{figure}

The MC is observed to be poleward at the surface with a maximum speed of $v_0 \simeq 20 ~\rm ms^{-1}$ at midlatitude \citep{2010ApJ...717..488B}, but helioseismic inversion are currently unable to probe its amplitude in layers much below $0.95 R_\odot$ as it possesses weaker flows that are difficult to detect \citep{2010Sci...327.1350H,1997SPD....28.1002G,2002ApJ...570..855H}. Nevertheless, this element is crucial in flux transport models incorporating such process because it links the two sources of the magnetic field, namely the BCZ and the solar surface. In theoretical models, one usually assumes to conserve mass using a unicellular flow with an equatorward return flow at the BCZ. This has been reproduced in recent solar simulations \citep{2008ApJ...673..557M} when global flows are averaged over long intervals. However, other simulations tend to show that it is instead rather multicellular both in radial and latitudinal direction and highly time dependant \citep{2002ApJ...570..865B,browning2006}.

The nature of these flows in other stars is even less constrained. In more rapidly rotating Suns, 3-D models presently indicate that the circulations are likely to be multi-celled in both radius and  latitude \citep{2007ApJ...669.1190B,brown2008}. In the series of models discussed in this paper, we restrict ourselves to Babcock-Leighton flux-transport models that have a large single cell per hemisphere. As in the Sun, the meridional circulation are directed poleward at the surface and here they vanish at the bottom boundary ($r=0.6$). This flow penetrates a little beneath the tachocline into the radiative interior as it is likely to occur \citep{2008ApJ...674..498G,brun2011}. To model the single cell meridional circulation we consider a stream function with the following expression from \citet{2011ApJ...733...90D} :

\begin{eqnarray}
  \psi &=& - \psi_0 \left(\frac{r-r_b}{1-r_b}\right)^2(\theta-\theta_0) \sin\left(\pi\frac{r-r_b}{1-r_b}\right)\\
  &&\times \rm e^{-\left(\frac{r-r_0}{\Gamma}\right)^2} (1-\rm e^{-\beta_1r\theta^\epsilon})(1-\rm e^{-\beta_2r\theta^\epsilon}) \nonumber
\end{eqnarray}

\noindent where $r_b=0.65$, $\theta_0=0$, $\Gamma=0.15$, $r_0=0.76$, $\epsilon=2.0$, $\beta_1=0.316228$, $\beta_2=0.3$. $\psi_0$ is chosen such that the velocities are normalised to unity. We define $\rho$ the density given by

\begin{equation}
  \rho(r) = \left[\frac{1}{r}-0.97\right]^m
\end{equation}

in which $m=1.5$ The velocity components are derived through the relation ${\rho \bf v_{\rm p}}=\nabla \times(\frac{\psi}{r \sin\theta} \hat {\bf e}_{\phi})$. In our simulation, radial resolution spans from $0.6$ to $1.0 R_\odot$ corresponding to a density contrast of $\Delta\rho \simeq 120$  

With this setup, the ratio between the maximal speed at the surface  and the BCZ is $v_{\rm surf}/v_{\rm BCZ} \simeq 109$. This flow further vanishes at $r_b$. We chose this profile because it allow us to control this ratio, which we found to be an important parameter for the ability of pumping to influence the magnetic period. The importance of pumping is enhanced as the MC amplitude is decreased at the BCZ, \ie there is a stronger correlation between the magnetic cycle period and the amplitude of pumping.

We turn now to the description of the turbulent pumping, the process we will focus on. In principle, its characteristics can be determined by direct numerical integration of the equations of magnetohydrodynamics (MHD). Several authors \citep[e.g.][]{1990A&amp;A...232..277B,2001A&amp;A...376..713O} have attacked this problem in order to calculate the dynamo coefficients, \ie the $\alpha$-effect and $\gamma$ -pumping. However, physical conditions in the solar convection zone (\eg high Reynolds number) prevent simulations from computing these coefficients in an entire shell. For this reason, most of 3D MHD simulations are restricted to Cartesian boxes that represents only a small section (in both radius and latitude) of stellar convection zone \citep[see however][for recent progress in computing transport coefficients in global model]{2010ApJ...711..424B}.  The profile we use in this work are inspired by one of the last estimation done by \citet{2006A&amp;A...455..401K}.

\begin{eqnarray}
  \gamma_r&=&-\frac{1}{4}\left[1+\erf{\frac{r-0.715}{0.015}}\right] \nonumber \\
  &&\times\left[1-\erf{\frac{r-1.02}{0.05}}\right] \nonumber\\
  &&\times \cos^2(\theta) \\
  \gamma_\theta&=&\frac{25\sqrt{5}}{64}\left[1+\erf{\frac{r-0.74}{0.03}}\right] \nonumber \\
  &&\times \left[1-\erf{\frac{r-0.94}{0.03}}\right] \nonumber \\
  &&\times \cos(\theta)\sin^4(\theta)
\end{eqnarray}

\begin{figure}[!h]
  \includegraphics[width=8cm]{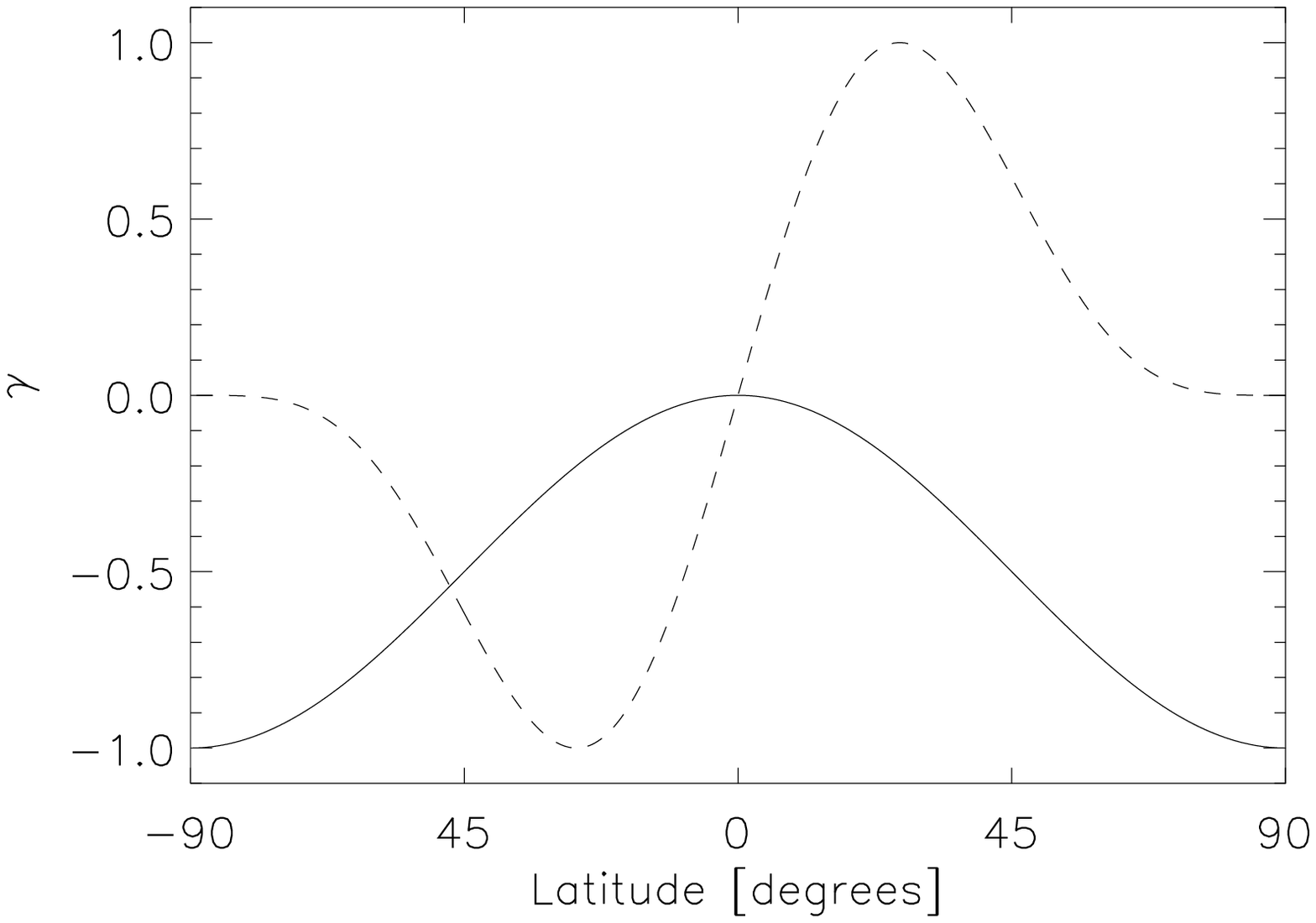}
  \includegraphics[width=8cm]{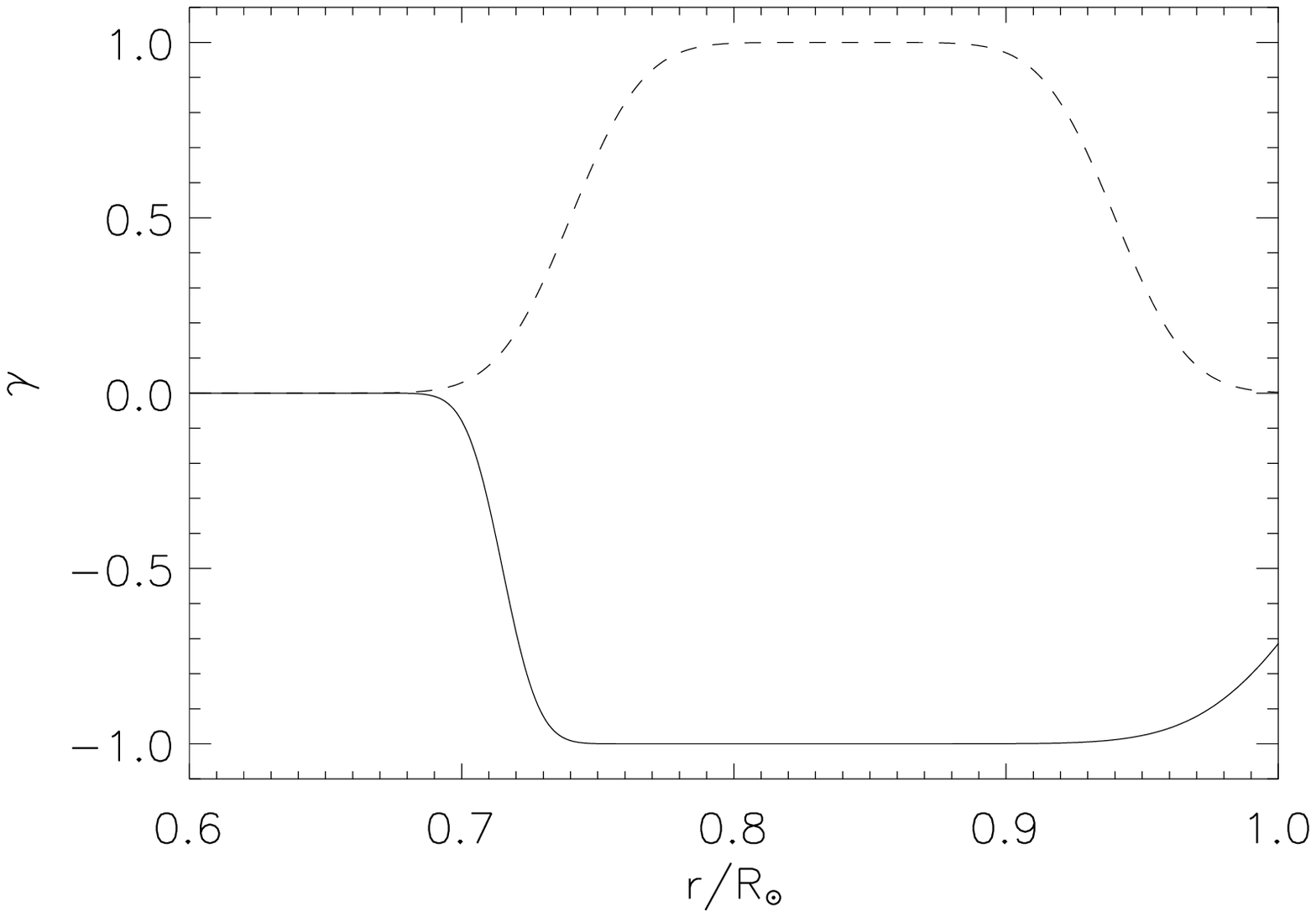}
  \caption{Profile of $\gamma_\theta$ (dash line) and $\gamma_r$ (solid line) as function of latitude at their maximal radial strength, \ie at r=0.85 for both cases (upper panel), and as function of radius at their maximal latitudinal strength, \ie at $\theta=0$ for $\gamma_r$ and $\theta=\acos(1/\sqrt{5})$ for $\gamma_\theta$ (lower panel).}
  \label{profile_gamma}
\end{figure}

\subsection{The standard model}\label{section_standardmodel}
We first compute a model \emph{without} the pumping effect which we will refer as our \emph{standard} case for direct comparison with forthcoming models. As we want to reproduce the Sun's properties, we have chosen our parameters for case $\bf{S}$ (see Table \ref{table_parameters}) such as to get the correct cycle period. We choose $C_s$ such that it is well above the threshold for dynamo action ($C_s^{\rm crit}=9.6$), enabling us to see the fields in the regime of a well established flux transport dynamo.

\begin{figure}[!h]
  \centering
  \includegraphics[width=8cm]{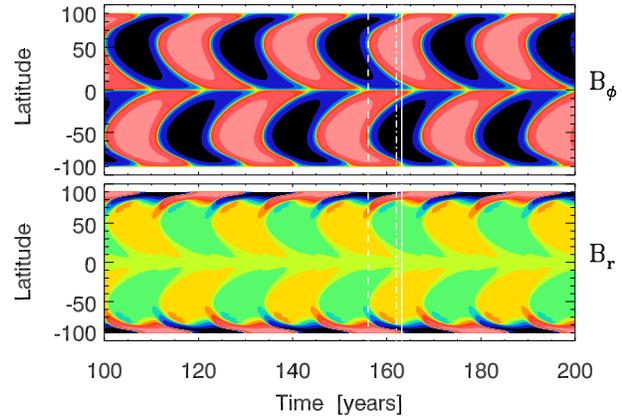}
  \caption{Butterfly diagram, \ie time-latitude slice of the toroidal field at the BCZ and of the radial field at the surface. Red (blue) colors indicate positive (negative) magnetic fields. Contours are logarithmically spaced with 2 contours covering a decade in field strength. The color range spans from Max to Min of the magnetic field strength. The vertical dashed line corresponds to the epoch of reversal of toroidal field, the plain line correspond to the epoch of reversal of poloidal field at the poles from negative to positive polarity and the dash-dotted line corresponds to the positive maximum of toroidal field near the equator.}
  \label{diagpap_standard}
\end{figure}

We show in Fig. \ref{diagpap_standard} the radial field at the surface and the toroidal field at the BCZ, \ie at $r=0.7 R_\odot$. It can be directly compared to the solar butterfly diagram if we identify the toroidal field at the BCZ as the source of buoyant magnetic flux tubes rising radially through the convection up to the surface seen as active regions. For this, the rising time must be very short compare to the magnetic cycle. But according to models, the rising time spans from months to dozens of years \citep[see review of ][]{lrsp-2009-4}. In particular, \citet{2010A&amp;A...519A..68J} found that even a short delay is important for a modulation in the cycle but not on the cycle period itself. We will thus not address this problem in this work.

With this typical model, we are able to reproduce several aspects of the solar cycle, notably its period of approximately 22 years, a strong equatorward branch for toroidal field restricted to low latitudes, a phase shift of $\pi/2$ between the surface polar field and the deep toroidal field, so that the polar field changes its polarity from negative to positive when the toroidal field is positive and maximal in intensity near the equator.

As described in Section \ref{section_ingredients}, the amplitude of MC is very low at the BCZ compare to the velocity at the surface. In the advection dominated regime, we found that this strongly affects the magnetic cycle periods because it is the dominant mechanism capable of transporting the toroidal field toward the equator. The recent work from \citet{2011ApJ...738..104P} confirm the great importance of the speed and the depth of the return flow.  To compensate for this effect and to recover the 22-year magnetic cycle, we choose a maximal velocity at the surface ($\sim 32 \ms$) slightly higher than what is usually assumed in simulations ($\sim 20 \ms$). However, meridional flow is a very time-dependant process which can reach values as high as what is used in this work \citep{2010ApJ...717..488B}. Also, the strong equatorward branch for the toroidal field is the signature of the drag of the toroidal field by equatorward MC at the BCZ and thus clearly shows the dominating effect of field advection over diffusion which can be seen in the scaling law :

\begin{equation}
  \Pcyc \propto \Omega_0^{0.05} s_0^{0.07} v_0^{-0.83}
\end{equation}

\noindent in the range $s_0=[4.7 ; 100] \cms$, $v_0=[1200;4500] \cms$ and $\Omega_0=[1.4 \times 10^{-6};7.2 \times10^{-6}]$ Hz. We directly see that the meridional circulation speed dominates over $C_s$ and $C_\Omega$. As $v_0$ dominates the scaling, it is crucial to assess how $v_0$ scales with $\Omega_0$. We rely on 3D numerical simulation from \citet{brown2008} which studies the influence of rotation rates on solar like stars. They found that MC decreases with the rotation rate as $v_0 \propto \Omega_0^{-0.45}$. This is not intuitive as one could expect that the meridional circulation increases with the rotation rate. A careful study of the vorticity equation shows that it actually weakens with rotation rate as more and more kinetic energy is being transferred to longitudinal motions at the expense of meridional kinetic energy. Assuming this relation in our simulations, we obtain naturally that $P_{\rm cyc} \propto \Omega_0^{0.41}$ in the range $\Omega_0=[1.0 \times 10^{-6}; 1.5 \times 10^{-5}] \rm ~Hz$. This is in agreement with what has been found in  JBB2010.

As stated above, the current model reproduce quite well a solar-like butterfly diagram, but is still unable to reproduce the $\Prot - \Pcyc$ relationship because of the strong influence of MC in these models, especially at the BCZ. To address this issue, JBB2010 tried to incorporate a multicellular MC as observed in 3D numerical simulations of \eg \citet{2007ApJ...669.1190B}  and \citet{brown2008}, and were indeed able to recover the observational trend. In this work, we propose another solution, that is to keep a unicellular MC but to introduce the turbulent pumping as a new mechanism to transport the poloidal field from the surface down to the tachocline. GdG2008 offered insights on the possibilities of such process for a shallow MC (see Section \ref{section_turbulentpumpingintro}). We wish here to verify if their results holds for a deeper MC and under which conditions the turbulent pumping can shorten the advection path driven by the meridional circulation.

\begin{table}[!h]
  \centering
  \caption{Summary of the control parameters defined in Section \ref{section_ingredients} for the different cases studied in this work. S and R stands for Standard and Reference, \ie they include (respectively dot not include) turbulent pumping. The trailing number in the case name gives the rotation rate $\Omega$ in units of $\Omega_\odot$. The S and R cases are at the solar rotation rate. The advective control parameters ($R_e$, $C_{\gamma r}$ and $C_{\gamma \theta}$ are normalized such that their relative strength can be directly estimated.}
  \label{table_parameters}
  \begin{tabular}{ccccccc} \hline
     {\small Case}    & $\frac{C_\Omega}{10^5}$ & $\frac{R_e}{10^3}$ & $\frac{C_{\gamma r}}{10^3}$& $\frac{C_{\gamma \theta}}{10^3}$ & $\scriptstyle\Pcyc$&$B_{\rm pol}/B_{\rm tor}$\\
    &               &    &    &      &{\small (years)} & \\
    \hline
    S   &  2.78   &  4.50   &  0      & 0 & 22.5 &          2.57          \\
    R0.7   & 2.00    &  2.67    &   0.0289    & 0.0723 & 26.4 & 1.51 \\
    R0.9   & 2.50    &  2.41    &   0.0452    & 0.113 & 24.2  & 0.999 \\
    R        &  2.78   &  2.30    &  0.0557      & 0.139  & 22.7  & 0.701 \\
    R1.1   &  3.00   &  2.22    &   0.0650    & 0.163 & 21.2  & 0.536\\
    R1.3   & 3.50    &  2.07    &   0.0885    & 0.221 & 18.7  & 0.328 \\
    R1.4   & 4.00    &  1.95    &   0.116    & 0.289 & 16.3   & 0.228 \\
    R1.6   & 4.50    &  1.85    &   0.146    & 0.366 & 14.5    & 0.174 \\
    R1.8   & 5.00    &  1.76    &   0.181    & 0.452 & 13.0   & 0.148 \\
    R2.0   & 5.50    &  1.69    &   0.219    & 0.547 & 11.7   & 0.127 \\
    R2.5   & 7.00    &  1.52    &   0.354    & 0.885 & 9.22   & 0.104 \\
    R2.9   & 8.00    &  1.43    &   0.462    & 1.16 & 8.22    & 0.100 \\
    R3.6   & 10.0    &  1.29    &   0.723    & 1.81 & 6.76     & 0.0706 \\
    R4.3   & 12.0    &  1.19    &   1.04    & 2.60 & 5.14      & 0.0539 \\
    R5.0   & 14.0    &  1.11    &   1.42    & 3.54 & 4.51       & 0.0377 \\
    \hline
  \end{tabular}
\end{table}

\section{Influence of turbulent pumping on stellar cycles}\label{section_pumping}
\subsection{Reference case}
\begin{figure}[!h]
  \centering
  \includegraphics[width=8cm]{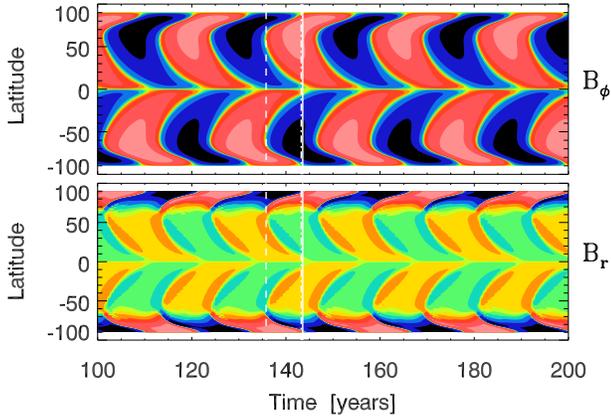}
  \caption{Butterfly diagram for the reference case. Same color code as in Fig. \ref{diagpap_standard}.}
  \label{diagpap_reference}
\end{figure}

We turn now to the models \emph{with} turbulent pumping. As before for the standard case, we define here the reference case $\bf{R}$ corresponding to the Sun, \ie at the solar rotation rate (cf Fig. \ref{diagpap_reference}) and reproducing the solar features described in Section \ref{section_standardmodel}. We rely on the estimation from \citet{2006A&amp;A...455..401K} for the pumping amplitude with $\gamma_{r 0}=40 \cms$ and $\gamma_{\theta 0} = 100 \cms$. We sum up the physical parameters for the reference case (\ie with turbulent pumping) at 1 solar rotation rate as R in Table \ref{table_parameters}.

With the profiles described in Section \ref{section_ingredients}, the latitudinal pumping component increases the total advective speed resulting in a lower MC amplitude ($\sim 17 \ms$) to keep a 22 year period, in a better agreement with temporally averaged observations. The strong equatorward branch of the toroidal field appears at slightly higher latitudes $65^\circ$ with the regions of strongest magnetic intensity confined in a smaller area near the poles. We have the opposite situation at the surface where the effective speed is reduced expanding the poloidal field to lower latitudes. At the poles, the radial pumping drags the poloidal field down to the tachocline and thus leads to a lower concentration of surface magnetic fields at the poles. This effect is in our deep MC not as striking as what has been reported by GdG2008. The phase relation of $\pi/2$ between the polar and the toroidal fields described in Section \ref{section_standardmodel} are well reproduced in this case (see Fig. \ref{diagpap_reference}).

\begin{figure}[!h]
  \centering
  \includegraphics[height=3.2cm,trim=8.2cm 4.6cm 5.cm 2.2cm,clip]{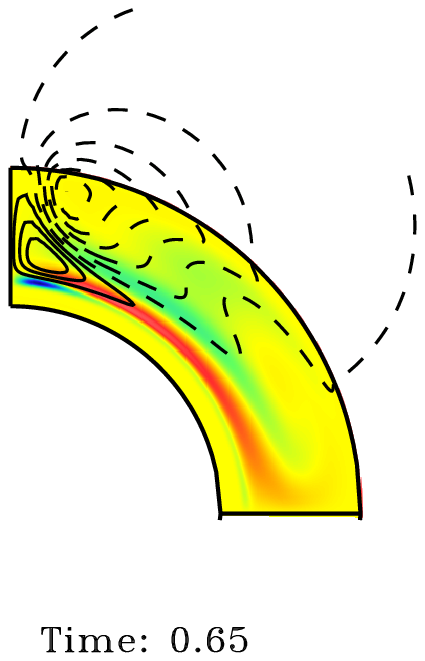}
  \includegraphics[height=3.2cm,trim=8.2cm 4.6cm 5.cm 2.2cm,clip]{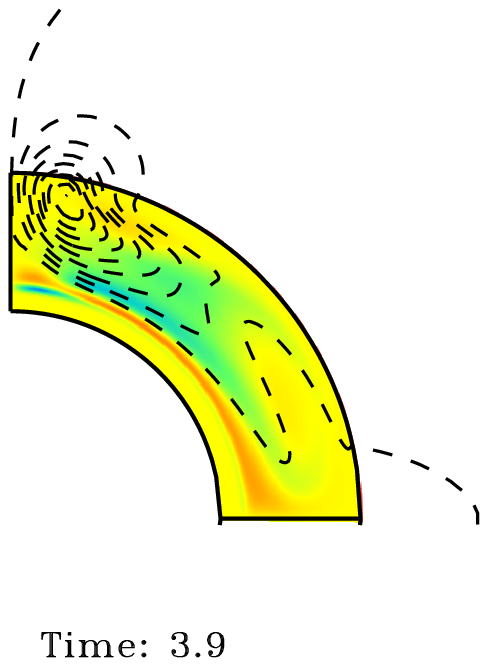}
  \includegraphics[height=3.2cm,trim=8.2cm 4.6cm 5.cm 2.2cm,clip]{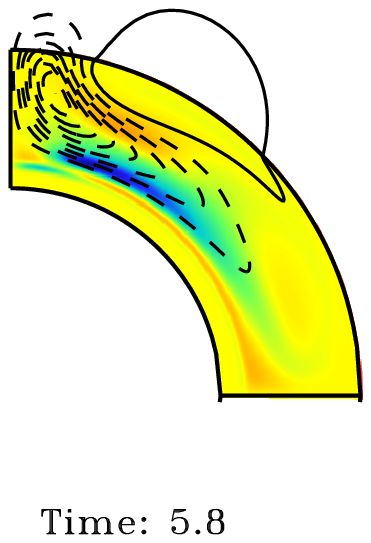}
  \includegraphics[height=3.2cm,trim=8.2cm 4.6cm 5.cm 2.2cm,clip]{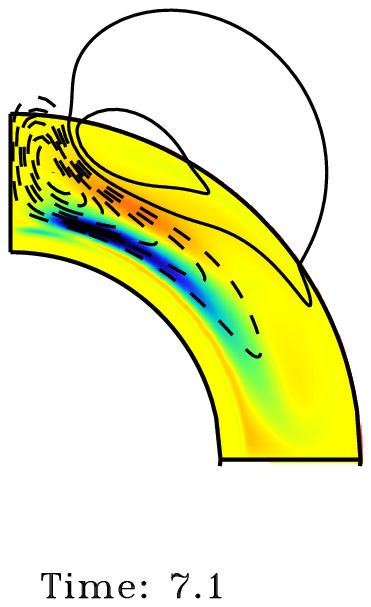}
  \includegraphics[height=3.2cm,trim=8.2cm 4.6cm 5.cm 2.2cm,clip]{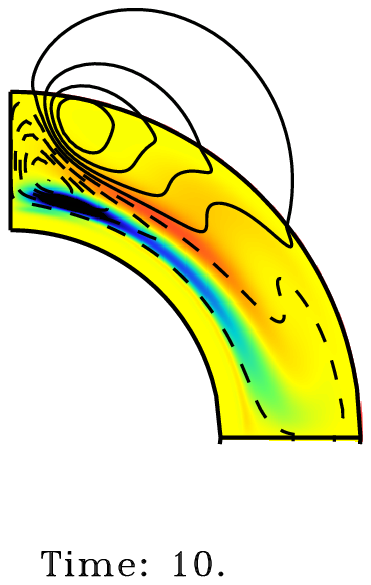}
  \includegraphics[height=3.2cm,trim=8.2cm 4.6cm 5.cm 2.2cm,clip]{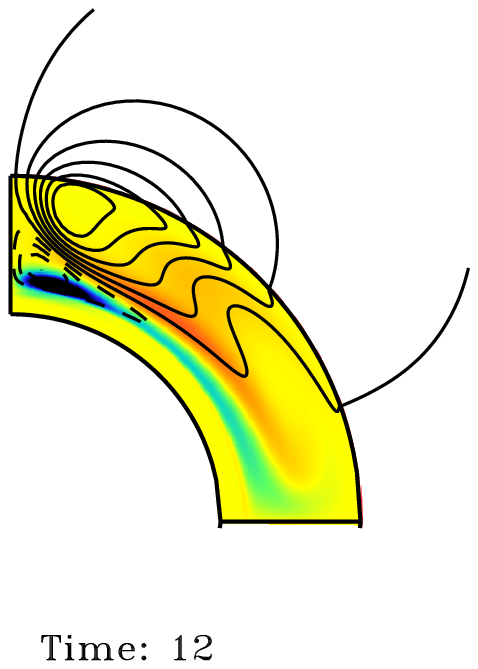}
  \caption{Evolution of the vector potential (black lines) and of the toroidal field (color contours) during half a magnetic cycle. Red (blue) colours indicate positive (negative) toroidal field and plain (dotted) lines indicate clockwise (anticlockwise) poloidal field lines. Time is expressed in years.}
  \label{fluxtransport_reference}
\end{figure}

Figure \ref{fluxtransport_reference} shows the behaviour of the magnetic field in this model. Here the magnetic field follows the advective path created by the meridional flow and consequently this large scale flow plays a key role in these dynamo solutions. In this reference case at $1\Omega_\odot$, pumping amplitudes are not sufficiently high to advect the magnetic fields away from MC flow.

\subsection{Varying the rotation rates}
We wonder now under which conditions the turbulent pumping can solve our initial problem, namely the scaling relationship between the magnetic cycle period $\Pcyc$ with the rotation period $\Prot$. A least square fit through our data gives

\begin{equation}
  P_{\rm cyc} \propto v_0^{-0.40} \gamma_{r0}^{-0.30} \gamma_{\theta 0}^{-0.15} \label{scaling_reference}
\end{equation}

\noindent in the range of $v_0=[1000;4500]$ $\rm cm ~s^{-1}$, $\gamma_{r0}=[20 ; 100]$ $\rm cm~s^{-1}$, $\gamma_{\theta 0}=[50 ; 300]$ $\rm cm~s^{-1}$.  This result is not in complete agreement with the work of GdG2008. We found that indeed the turbulent pumping becomes a major player in setting the magnetic period, but the quantification of its influence remains different. First, the MC is still the dominant effect and the radial pumping component is not as important as in GdG2008. Second, the effect of $\gamma_\theta$ is not negligible. This supports the idea that the latitudinal advection process, and especially at the BCZ, is an important ingredient in advection dominated BL models, capable of transporting the toroidal magnetic field from the pole toward the equator. This difference may come from their choice of a shallow MC with almost zero velocity at the BCZ.

A simple look at the scaling law \ref{scaling_reference} gives that if we want to recover the observational trend (again, we assume that $v_0 \propto \Omega^{-0.45}$), and assuming that $\gamma_r/\gamma_\theta$ remains constant, the pumping effect should roughly scales as $\Omega_0^2$. A first estimate done in \citet{2001ApJ...549.1183T} have shown that the turbulent pumping actually decreases with rotation rate. Later on, \citet{2009A&amp;A...500..633K} found on the contrary that the rotation rate have almost no effect on the vertical pumping. This lets the scaling as an open question. In order to verify if this scaling holds in solar type stars, work are currently done in 3D MHD simulation \citep{matt2011}.

\begin{figure}[!h]
  \centering
  \includegraphics[width=8cm]{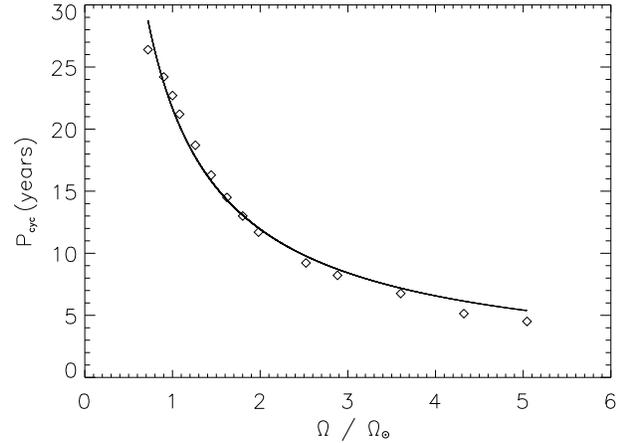}
  \caption{Magnetic cycle period as function of the rotation rate in models including turbulent pumping. Solid line is a least square fit of the simulated data.}
  \label{scaling_pcyc_omega}
\end{figure}

\begin{figure}[!h]
  \centering
  \includegraphics[width=8cm]{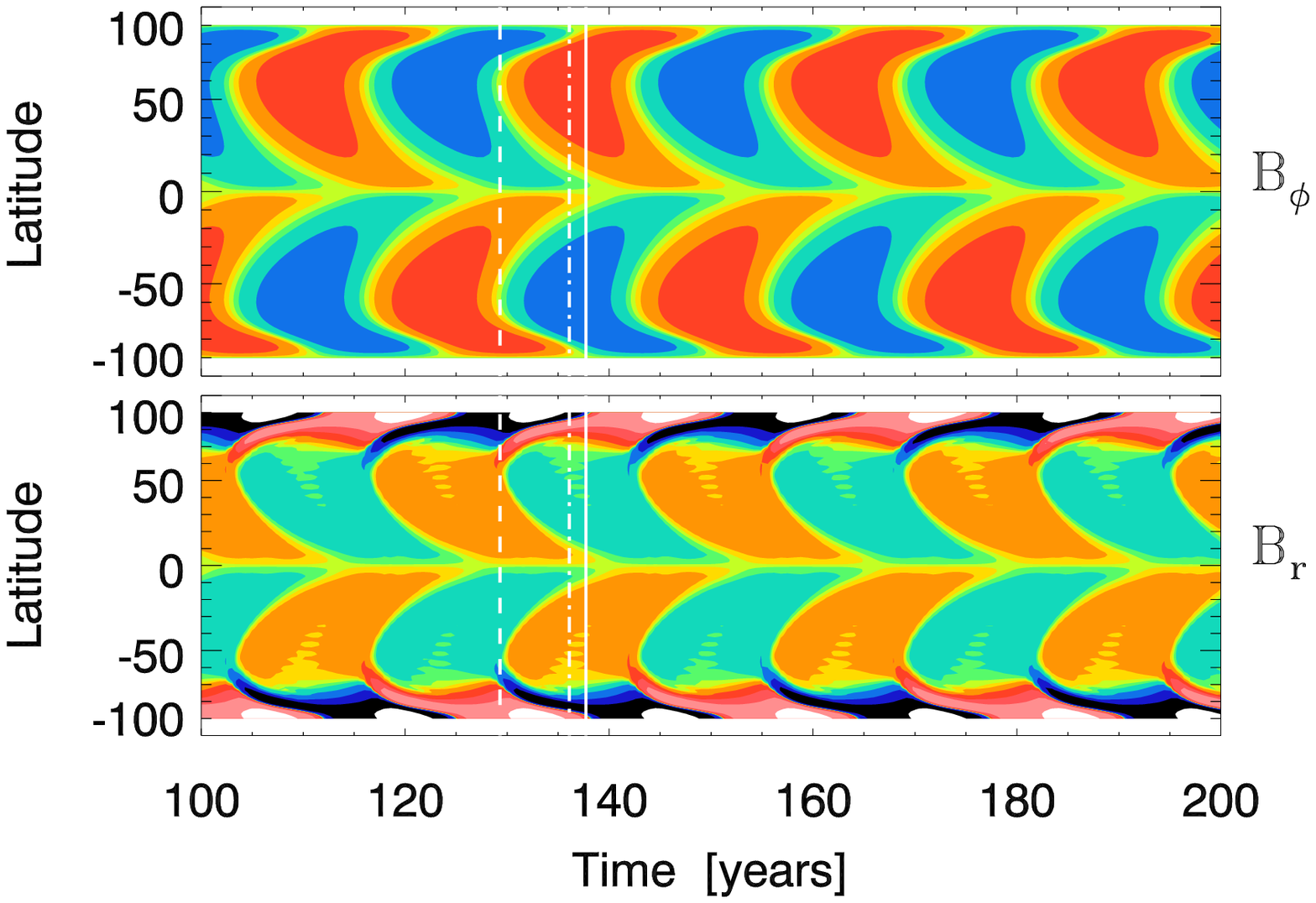} 
  \includegraphics[width=8cm]{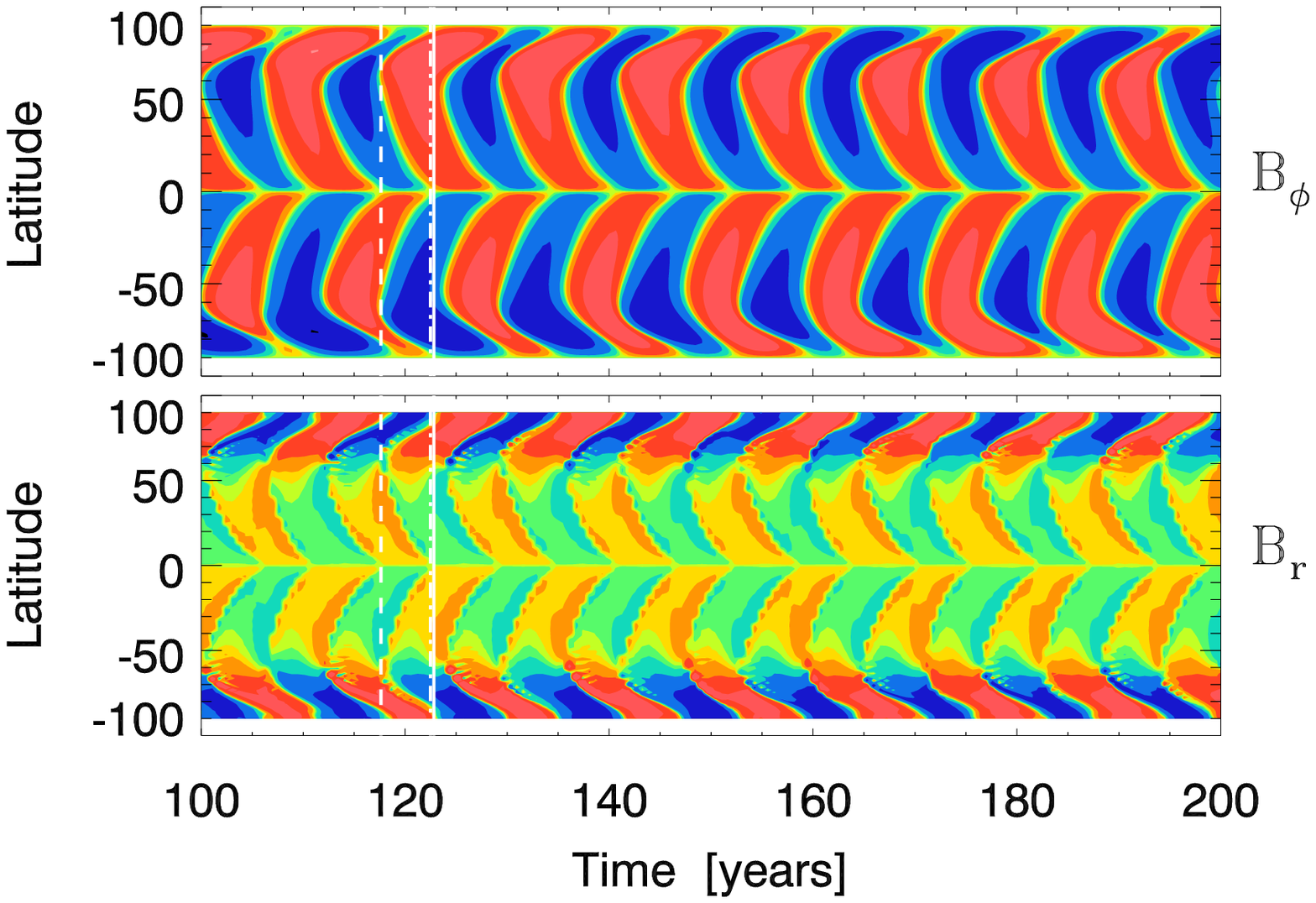} 
  \includegraphics[width=8cm]{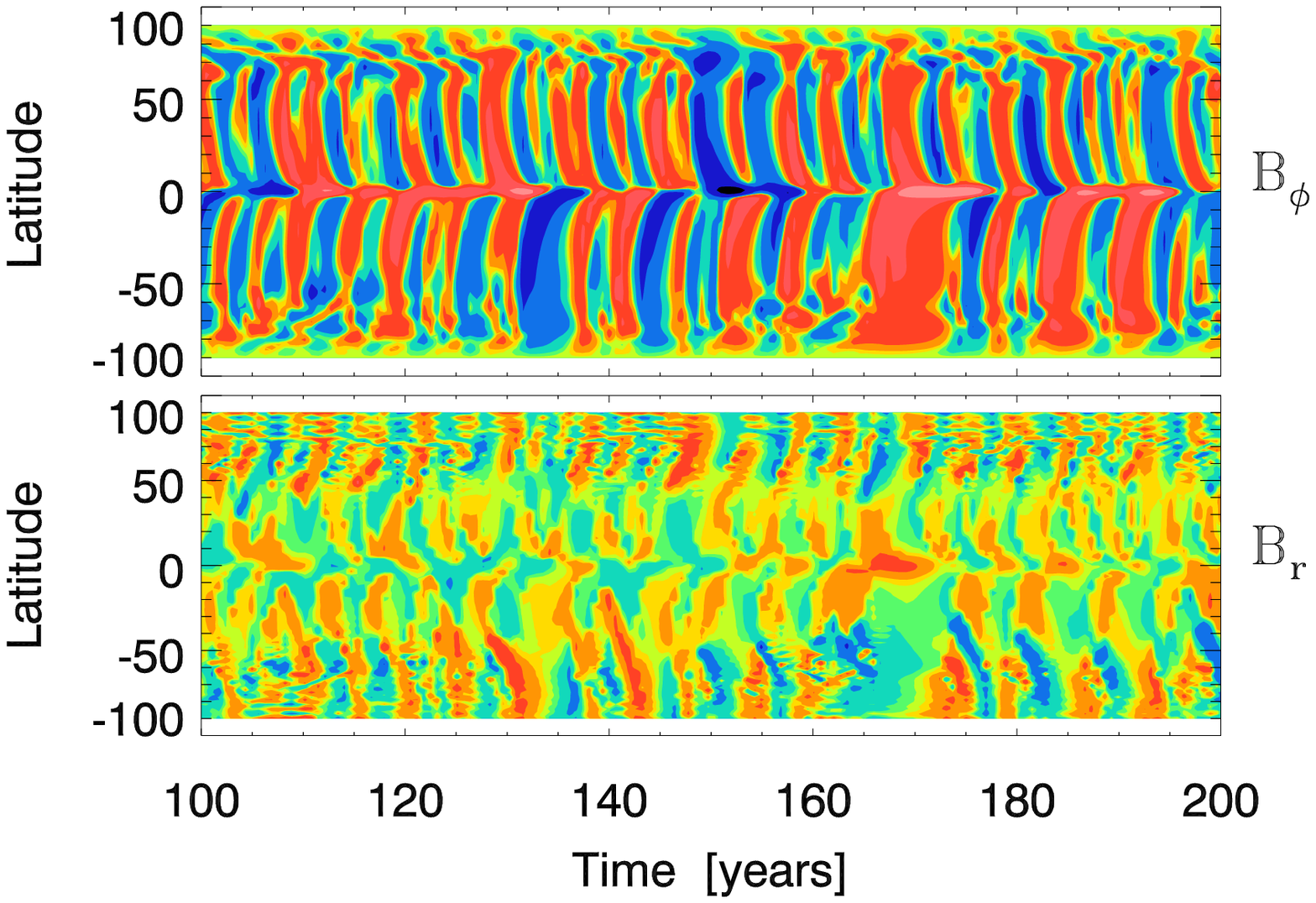} 
  \caption{Butterfly diagram for 3 representative cases : $0.7 \Omega_\odot$, $2.0\Omega_\odot$ and $5\Omega_\odot$. Same color code as in Fig. \ref{diagpap_standard}. All of the figures share the same color scale : between $-5 \times 10^3 \rm G$ and $5 \times 10^3 \rm G$ for $B_r$ and between $-9 \times 10^5 \rm G$ and $9 \times 10^5 \rm G$ for $B_\phi$. We show the phase relations for the top and middle panel only as it is not possible to define them for the highest rotation case.}
  \label{diagpap_rotation}
\end{figure}

Under these assumptions, we found indeed that $P_{\rm cyc} \propto \Omega_0^{-0.86}$, in agreement with the observations as expected. On Fig. \ref{scaling_pcyc_omega}, we present the rotation rates ranging from $0.7\Omega_\odot$ up to $5 \Omega_\odot$ with a least square fit of the data. Our result does not hold outside this range because a systematic period does not emerge for $\Omega > 5 \Omega_\odot$. For $\Omega < 0.7 \Omega_\odot$, the pumping is negligible and lets MC imposing that $\Pcyc$ increases with $\Omega$. We are thus back to a strong dependency of $\Pcyc$ with MC amplitude. We show in Fig. \ref{diagpap_rotation}, 3 representative cases at $0.7 \Omega_\odot$, $2\Omega_\odot$ and $5 \Omega_\odot$. We see that the equatorward branch becomes shorter and shorter as the rotation rate, and so the pumping, is increased.  Also, the surface magnetic field becomes homogeneously distributed in latitude thanks to the increase of both components of the turbulent pumping. However, the determination of a period for the most rapidly rotating stars ($\Omega > 4 \Omega_\odot$) becomes difficult as the butterfly diagram is affected by more and more intermittency and small scale structures as can be seen in the bottom panel of Fig. \ref{diagpap_rotation}. This explains the last points of Fig. \ref{scaling_pcyc_omega} to lie slightly away from the trend.

One can also note the appearance of a modulation of both cycle strength and period as soon as $2\Omega_\odot$ as seen in the middle panel of Fig. \ref{diagpap_rotation}. We might not compare this with long term modulation known as Gleissberg cycles. For 2D models in the kinematic regime (as in this work), the Lorentz force exerted by the magnetic fields on the velocity field, the so-called Malkus-Proctor effect \citep{1975JFM....67..417M}, is not included. Such a feedback from the large scale Lorentz force has been shown to have strong effect on for instance torsional oscillations \citep{2006ApJ...647..662R}, long term modulation \citep{2002A&amp;A...392..713P} and intermittency \citep{2000MNRAS.315..521M}. In the current simulations instead, the modulations are due to the pumping mechanism which advects the magnetic fields against the MC flow. Thus, the phase becomes more and more modified between $B_{\rm pol}$ and $B_{\rm tor}$ up to a point where ephemeral dynamo loops appears, generating multiple periodicities. It becomes even more dramatic when we keep increasing the rotation rate. The solar butterfly diagram features almost vanish in very small structures. The poleward and equatorward branches barely exist and the magnetic field is  completely homogeneously distributed over all latitudes and do not present a strong concentration of poloidal fields at the poles. In these cases, the advection is no longer dominated by MC but by pumping, suggering that we are entering in a new class of regime for BL flux transport models.

\begin{figure}[!h]
  \centering
  \includegraphics[width=8cm]{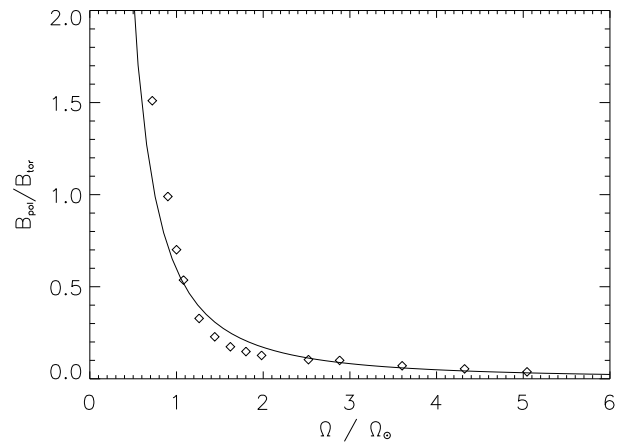}
  \caption{$B_{\rm pol}/B_{\rm tor}$ ratio as function of the rotation rate. Solid line is a least square fit of the data.}
  \label{bpolbtor_omega}
\end{figure}

We turn now to the ratio between the maximal value of the poloidal field at the surface and the maximal value of the toroidal field at the BCZ, $B_{\rm pol}/B_{\rm tor}$. This ratio is found to decrease with the rotation rate ($B_{\rm pol}/B_{\rm tor} \propto \Omega^{-1.80}$) as seen on Fig. \ref{bpolbtor_omega}. This is in reasonable agreement with the observations of rapidly rotating solar like stars by \citet{2008MNRAS.388...80P} where rapid rotators host a large scale toroidal component in their surface field whereas the magnetic field is mostly poloidal for low rotation rates. In the range of available observations, we found systematically a lower ratio. This is not surprising as we estimated $B_{\rm tor}$ at the BCZ where it is generated, and hence where it is the strongest whereas observers have access only at the weaker surface toroidal field. To evaluate the surface toroidal magnetic field in our simulations, a first approach would be to take the value close to the surface. A typical ratio between the BCZ ($r=0.7$) and the surface ($r=0.98$) toroidal fieldfor the reference case is found to be $B_{\rm tor}^{\rm BCZ}/B_{\rm tor}^{\rm surf} \simeq 60$.

\section{Conclusion and perspectives}\label{section_conclusion}
The aim of this study was to address the current issue encountered by 2D mean field models in which the magnetic period decreases with the rotation rate, contrary to the observations \citep{2009ASPC..416..375S,2011arXiv1109.4634W}. Earlier work of JBB2010 showed that this behaviour was due to the large influence of the MC. More specifically, they have used the results of recent 3D simulations \citet{brown2008} in which meridional circulation amplitude decrease with rotation rate, and came to the conclusion that the observed $\Pcyc - \Prot$ relationship cannot be reproduced, unless a multicellular MC is considered.

The idea of this work was to tackle this problem under another angle introducing the turbulent pumping mechanism which has been shown to have interesting properties on the magnetic period (GdG2008). We have then performed 2D BL flux transport simulations with the STELEM code in the advection dominated regime. We first computed a \emph{standard} model without turbulent pumping but producing solar characteristics. We found that in such models the magnetic period is indeed very sensitive to the MC amplitude ($\Pcyc \propto \Omega_0^{0.05} s_0^{0.07} v_0^{-0.83}$), and is therefore not able to reproduce the observations, confirming the results of JBB2010.

In the presence of turbulent pumping however, the MC is no longer the only process capable of influencing $\Pcyc$. We found that the latitudinal speed at the tachocline drives the magnetic cycle period. The weaker the MC, the stronger the influence of pumping on the magnetic cycle period. A reasonable equatorward flow at the BCZ ($\sim 0.2 \ms$) with a velocity contrast between the surface and the tachocline of $\sim 109$ gives the scaling $ P_{\rm cyc} \propto v_0^{-0.40} \gamma_{r0}^{-0.30} \gamma_{\theta 0}^{-0.15} $. The observational trend can be thus recovered only if $\gamma \propto \Omega^2$ which has not been reported yet in previous Cartesian simulations \citep{2001ApJ...549.1183T,2009A&amp;A...500..633K}. However 3D MHD simulations in full spherical shells are currently under development to test this assumption \citep{matt2011}.

Another successful feature resulting from the presence of turbulent pumping is their ability to reduce the strong concentration at the poles. On one hand, the strong radial component drags the surface field down to the tachocline. On the other hand, the latitudinal component is equatorward everywhere (and therefore at the surface) expanding the strong poloidal field to lower latitudes. As we increase the rotation rates, the pumping becomes stronger and this dilution is enhanced until the surface field becomes completely homogeneous. However, at this level of rotation, the butterfly diagram is strongly affected by the pumping which has become the dominant advective process with a greater amplitude than the MC. For instance, it does not show strong equatorward nor poleward branches. We enter here in a different dynamo regime, in which we are dominated , not by MC, but by turbulent pumping, with characteristics different than the Sun.

Although this new ingredient looks promising with respect to our study, turbulent pumping amplitude must be as high as few $\sim 10 \ms$ for $\sim 5 \Omega_\odot$. Such high values have not been reported yet in direct numerical simulations.  A way out would be to mix the two proposed approaches by considering in the same time both pumping and multicellular MC. The latter naturally arised in 3D MHD global simulations \citep{2002ApJ...570..865B,2007ApJ...669.1190B,brown2008} in which they last for several years and are supported by observational evidences via local helioseismology techniques \citep{2002AAS...200.0414H}.

Finally, long term modulations have not been considered in this work since no nonlinear effects beside quenching has been used. We intend to take into account those effects such as Malkus Proctor term in a future work since they are potentially important \citep{2000MNRAS.315..521M}. We also intend to extend the resolution domain by including an atmosphere which would give a prediction for the external toroidal field, but also a bottom boundary condition for stellar wind models \citep{2011ApJ...737...72P}.

Studying stellar magnetism with mean field models is instructive as we can extract information on the sensitivity of magnetic cycles to parameter change, such a study being very computationally expensive and delicate with 3D MHD simulations \citep{2011ApJ...731...69B}. However, in the light of this work, a subtle combination of different processes must be considered to account for the many aspects of the observations. More accurate data would be of great help to understand the complex underlying physics of stellar dynamo.

\acknowledgements
The authors acknowledge funding by the European Research Council through ERC grant STARS2 207430 \url{www.stars2.eu}. O. Do Cao  also acknowledges useful discussions with L. Jouve and G. Guerrero.

\bibliographystyle{astroads}
\bibliography{docao2011}

\end{document}